\documentclass[]{spie}  %>>> use for US letter paper
\usepackage[utf8]{inputenc}
\usepackage[british]{babel}
\usepackage{amssymb}

\usepackage{graphicx,rotating}
\graphicspath{{./graphics/}}  % look for figure graphics in this directory
\usepackage[colorlinks=true,allcolors=blue]{hyperref}

\usepackage[margin=2em,font=small,labelfont=bf]{caption}
\usepackage[capitalise]{cleveref}

\usepackage[T1]{fontenc}

\usepackage{textcomp,gensymb}

\usepackage{xspace}
\newcommand{\micron}{\ensuremath{\micro \mathrm{m}}\xspace}

\title{Practical Beam Transport for PFI \footnote{Copyright 2016 Society of Photo-Optical Instrumentation Engineers. One print or electronic copy may be made for personal use only. Systematic reproduction and distribution, duplication of any material in this paper for a fee or for commercial purposes, or modification of the content of the paper are prohibited. DOI: http://dx.doi.org/10.1117/12.2232382}}
\author[a]{David Mozurkewich}
\affil[a]{Seabrook Engineering, Seabrook, United States}
\author[b]{John Young}
\affil[b]{University of Cambridge, Cambridge, United Kingdom}
\author[c]{Michael Ireland}
\affil[c]{Australian National University, Mount Stromlo Observatory, Australia}

\authorinfo{Further author information: (Send correspondence to J.S.Y.)\\J.S.Y.: E-mail: jsy1001@cam.ac.uk}

\begin{document}
\maketitle

% :TODO: use consistent symbols

\begin{abstract}
  The Planet Formation Imager (PFI) is a future kilometric-baseline infrared interferometer to image the complex physical processes of planet formation. Technologies that could be used to transport starlight to a central beam-combining laboratory in PFI include free-space propagation in air or vacuum, and optical fibres. This paper addresses the design and cost issues associated with free-space propagation in vacuum pipes. The signal losses due to diffraction over long differential paths are evaluated, and conceptual beam transport designs employing pupil management to ameliorate these losses are presented and discussed.
\end{abstract}


\section{Introduction}
\label{sec:intro}
The Planet Formation Imager (PFI)~\cite{2014SPIE.9146E..10M} is a future interferometric facility to image the process of planet formation. The goal of PFI is to deliver the capability to image complex scenes at the resolution of a forming giant planet's Hill sphere, i.e.\ 0.5 milli-arcsec for a Jupiter-mass protoplanet at 1\,AU separation in a nearby star-forming region at 140\,pc. PFI will operate at mid-infrared wavelengths where the contrast between the protoplanet and the central star is maximized~\cite{Kraus_et_al-2016-PFI_science}, in either or both of the 3--6\,\micron ($L$ and $M$ bands) and 8--13\,\micron ($N$ band) ranges, implying a maximum baseline of order 5\,km. Both heterodyne~\cite{2014SPIE.9146E..12I} and homodyne~\cite{Ireland_et_al-2016-Status_PFI_concept} concepts are being considered for mid-infrared interferometry with PFI. Both architectures will require fringe tracking at near-IR wavelengths using homodyne beam combination (direct detection).

This paper will consider how to implement beam transport and optical delay lines for the near-IR and mid-IR wavebands, noting that physical delay lines are not required for heterodyne mid-IR interferometry. In developing the technology roadmap for PFI we are considering three possible solutions for beam propagation: optical fibres, vacuum pipes and free air propagation with adaptive optics correction.

Optical fibers are a potentially attractive solution, but multiple fiber designs will be needed to cover all of the wavebands of interest. To date, there has been little demonstrated success using fibers for beam transport in infrared interferometry. The design issues that will need to be addressed for fibers include throughput, path length stabilization, and dispersion. Alternatively, rather than propagate light beams in vacuum, it may be possible to use propagation in air with a suitable adaptive optics system to correct the resulting aberrations. This option has barely been studied by previous workers and would be the highest-risk choice.

In this paper we will focus on solutions using vacuum pipes. This option has the lowest technical risk, having been proven at existing arrays such as CHARA, probably the best performance, but may also have the highest cost. The need to install long runs of straight vacuum pipes will also drive the choice of suitable sites for PFI.

At least the following issues must be considered when designing beam transport and delay line solutions based on propagation in vacuum pipes:
\begin{itemize}
  \item The extent to which pupil relay can reduce the diffraction losses and hence allow the use of affordable-size vacuum pipes.

  \item The practical limits to the length of a continuously-variable delay line, for example the speed of repositioning.

  \item The performance impacts of using many passes through the delay lines, such as increased optical path length jitter.

  %\item Installation and maintenance cost for delay line vacuum pipe as a function of pipe length and diameter (assuming precision rails are not needed).

  \item The viability of design concepts for low-loss switchable fixed delays incorporating pupil relay, to allow the use of shorter variable delay lines.
\end{itemize}

In this paper we evaluate the signal loss for collimated beams due to diffraction (\cref{sec:diffraction}), and hence calculate the beam sizes needed for the heterodyne and direct detection architectures. An analysis of simple pupil management schemes is presented in \cref{sec:pupil-relay}. In \cref{sec:dl-concepts} concept designs for delay lines based on these schemes are introduced and discussed, including both switchable fixed delays and continuously-variable delay lines.

\section{Diffraction of collimated beams}
\label{sec:diffraction}

For this study, diffraction losses were calculated using numerical integration in two dimensions. The problem has azimuthal symmetry, so we can let $E_0(\rho)$ denote the initial electric field and $E(r)$ the field after it has propagated a distance $z$. Then, with $k=2\pi/\lambda$,
\begin{equation}
  E(r) = \left( \frac{k}{iz} \right) e^{ik \left( z + \frac{r^2}{2z} \right)} \int_{0}^{\rho_0} \rho E_0(\rho) e^{ik\rho^2/2z} J_0(kr\rho/z) d\!\rho
\end{equation}

Results can be scaled to other wavelengths, beam diameters and propagation distances by noting that these parameters enter the calculation only through the dimensionless Fresnel number
\begin{equation} \label{eq:fresnel-no}
  F = \frac{D^2}{\lambda z} \, .
\end{equation}
Given two beam profiles $a$ and $b$, the squared visibility is determined using a standard four-bin (``ABCD'') estimator:
\begin{equation}
  |V|^2 = \frac{(B_3 - B_1)^2 + (B_2 - B_0)^2}{(I_a + I_b)^2} \, ,
\end{equation}
with the coherent fluxes $B_k$ and incoherent fluxes $I_a$ and $I_b$ defined as follows:
\begin{equation}
  B_k = \int 2 \pi r \Re{E_a^* E_b e^{ik\pi/2}} dr \, ,
\end{equation}
\begin{equation}
  I_a = \int 2 \pi r |E_a|^2 dr \, ,
\end{equation}
\begin{equation}
  I_b = \int 2 \pi r |E_b|^2 dr \, .
\end{equation}

The calculated signal loss factor as a function of differential propagation distance $\Delta z$ is plotted in \cref{fig:propagation}, for various values of the shorter of the two propagation distances $z_a$ and $z_b$. The graph shows that the signal loss mostly depends on $\Delta z$ and is only weakly dependent on the absolute distance propagated.

\subsection{Application to PFI}

\begin{figure}
  \begin{center}
    \includegraphics[width=0.49\linewidth]{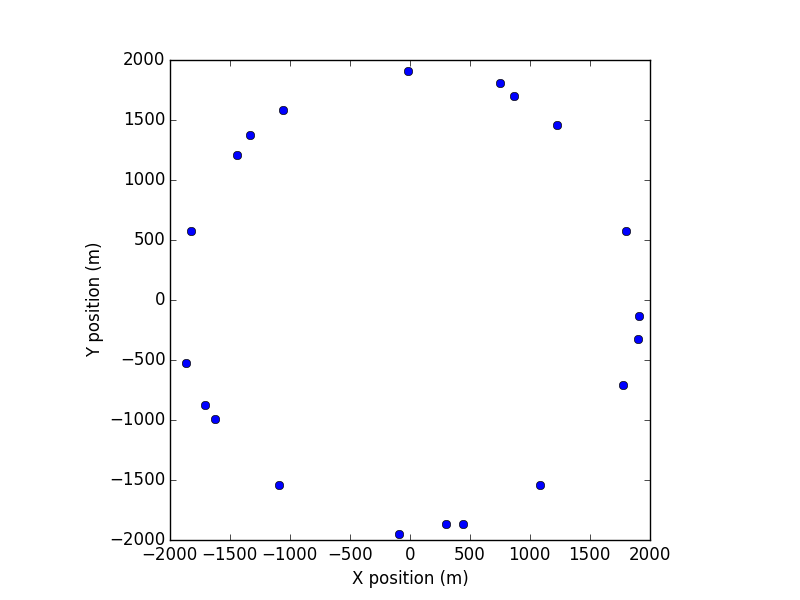}
    \includegraphics[width=0.49\linewidth]{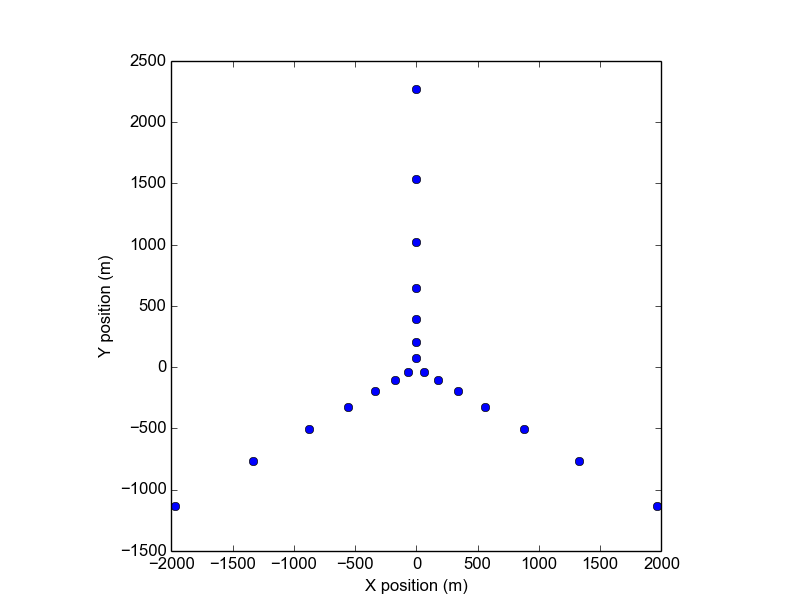}
  \end{center}
  \caption{\label{fig:arrays} Reference array layouts for PFI: a circular array with random perturbations (left), and a conventional Y-shaped array (right). Both arrays comprise 20 telescopes and have a maximum baseline of approximately 4\,km. These layouts have second-nearest-neighbor baselines shorter than 2\,km to ensure reliable self-referenced fringe tracking.}
\end{figure}

\begin{figure}
  \begin{center}
    \includegraphics[width=0.3\linewidth]{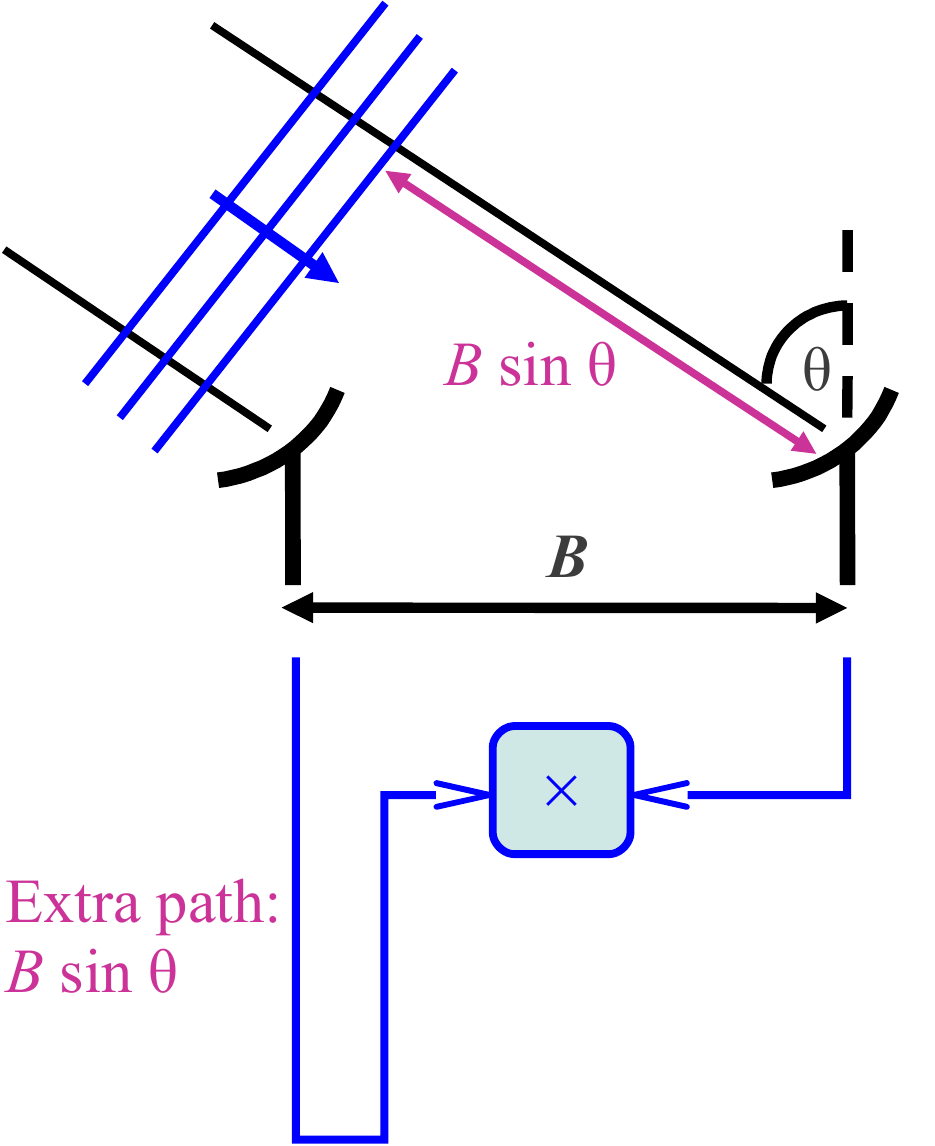}
  \end{center}
  \caption{\label{fig:geometric-delay} Diagram of a two-element interferometer receiving plane wavefronts from a distant source, showing the geometric delay that must be compensated using the interferometer's delay line.}
\end{figure}

Two reference designs for the array layout of PFI are shown in \cref{fig:arrays}. To evaluate the worst case propagation distances for each array, we must consider both propagation from the unit telescopes to the beam combining laboratory (assumed to be at the center of the array) and compensation of the geometric delay (\cref{fig:geometric-delay}) due to projection of the baseline. The maximum geometric delay is for a target on the horizon, aligned with the longest baseline. Of the two possible target directions, the one with the beam combining laboratory nearest to the target is the worst case --- the parameters for this case are presented in \cref{tab:arrays}.

\begin{figure}
  \begin{center}
    \includegraphics[width=0.75\linewidth]{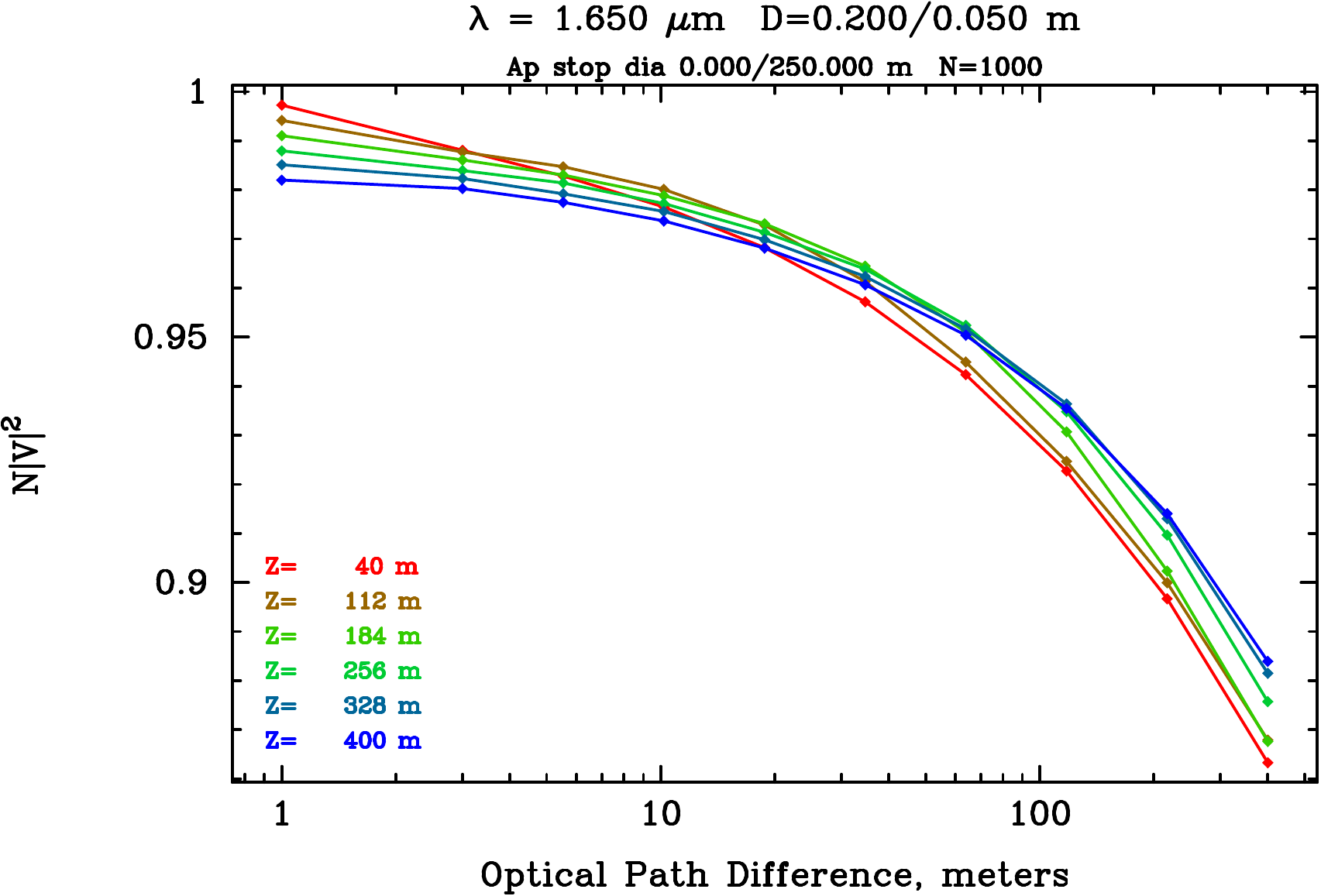}
  \end{center}
  \caption{\label{fig:propagation} Interferometric signal loss $N|V|^2$ due to differential propagation of collimated beams.  The various curves are for different absolute propagation distances $z$. The initial beam diameter is 0.2\,m with a 0.05\,m central obscuration, and the wavelength is 1.65\,\micron. The results may be scaled to other distances and wavelengths using Eq.~\ref{eq:fresnel-no}.}
\end{figure}

Scaling the results in \cref{fig:propagation} to $\Delta z$ of 4\,km (\cref{tab:arrays}) and the longest possible science wavelength of 13\,\micron, we find that a beam size of $D_0 \sqrt{\frac{z}{z_0}\frac{\lambda}{\lambda_0}} = 1.8$\,m would be needed to keep the signal loss due to diffraction to 10\% or less. If we restrict mid-IR operation to the $L$ (3.5\,\micron) and $M$ (5\,\micron) astronomical bands, the required beam size is reduced to 1.1\,m. Even if we decide to tolerate higher losses (say 20\%), and take advantage of the fact that diffraction can provide a helpful spatial filtering effect when atmospheric aberrations are present \cite{Horton_et_al-2001-Diffraction_losses}, it is clear that the cost of the vacuum pipes and supports to accommodate such large beams would be prohibitively expensive (both in absolute terms and in comparison with the cost of the unit telescopes).

\begin{table}
  \caption{\label{tab:arrays} Parameters of the reference pseudo-circular and Y-shaped arrays shown in \cref{fig:arrays}.  The worst-case (i.e.\ largest differential) propagation distances assume a target on the horizon and account for propagation from the telescopes to the array center followed by compensation of the geometric delay.}
  \begin{center}
    \begin{tabular}{lrr}
      \hline
      Parameter & Pseudo-circular & Y-shaped \\
                & array           & array    \\
      \hline
      $B_{\mathrm{min}}$ /m   &  163 &  130 \\
      $B_{\mathrm{max}}$ /m   & 3829 & 3935 \\
      Worst-case $Z_a$ /m & 1925 & 2272 \\
      Worst-case $Z_b = ( Z_a + B_{\mathrm{max}} )$ /m & 5761 & 6207 \\
      \hline
    \end{tabular}
  \end{center}
\end{table}

On the other hand, if a heterodyne solution were adopted, the beam size needed for the 1.65\,\micron fringe-tracking light, assuming a maximum $\Delta z$ of 2\,km for the nearest and second-nearest-neighbor baselines that will be used to co-phase the array, would be only 0.45\,m. This beam size could be accommodated in vacuum pipes with $\sim$0.7\,m inner diameter, for a feasible cost.

An interesting question is the relative costs of the beam transport hardware for the pseudo-circular and Y-shaped arrays. Both have similar worst-case differential propagation distances, but the beams from the telescopes on each arm travel in parallel to the central laboratory, allowing pipe supports and/or vacuum pipes to be shared between multiple beams. This aspect of the PFI cost model will be examined in more detail later in the design process.

\section{Pupil management}
\label{sec:pupil-relay}

There are several options for incorporating pupil management into the beam transport system. The delay lines could consist of either a combination of switchable fixed delays (``long delay lines'') and relatively short ($\lesssim 1$\,km) continuously-variable delay lines (as at NPOI and CHARA), or a long variable delay line (as adopted at MROI). For the first option, incorporating pupil management into the switchable fixed delays but not the variable delay lines might be adequate. If long continuously-variable delay lines are used, these would have to incorporate pupil management, for example using variable curvature mirrors as at VLTI \cite{2000SPIE.4006..104F}.

\begin{figure}
  \begin{center}
    \includegraphics[width=0.7\linewidth]{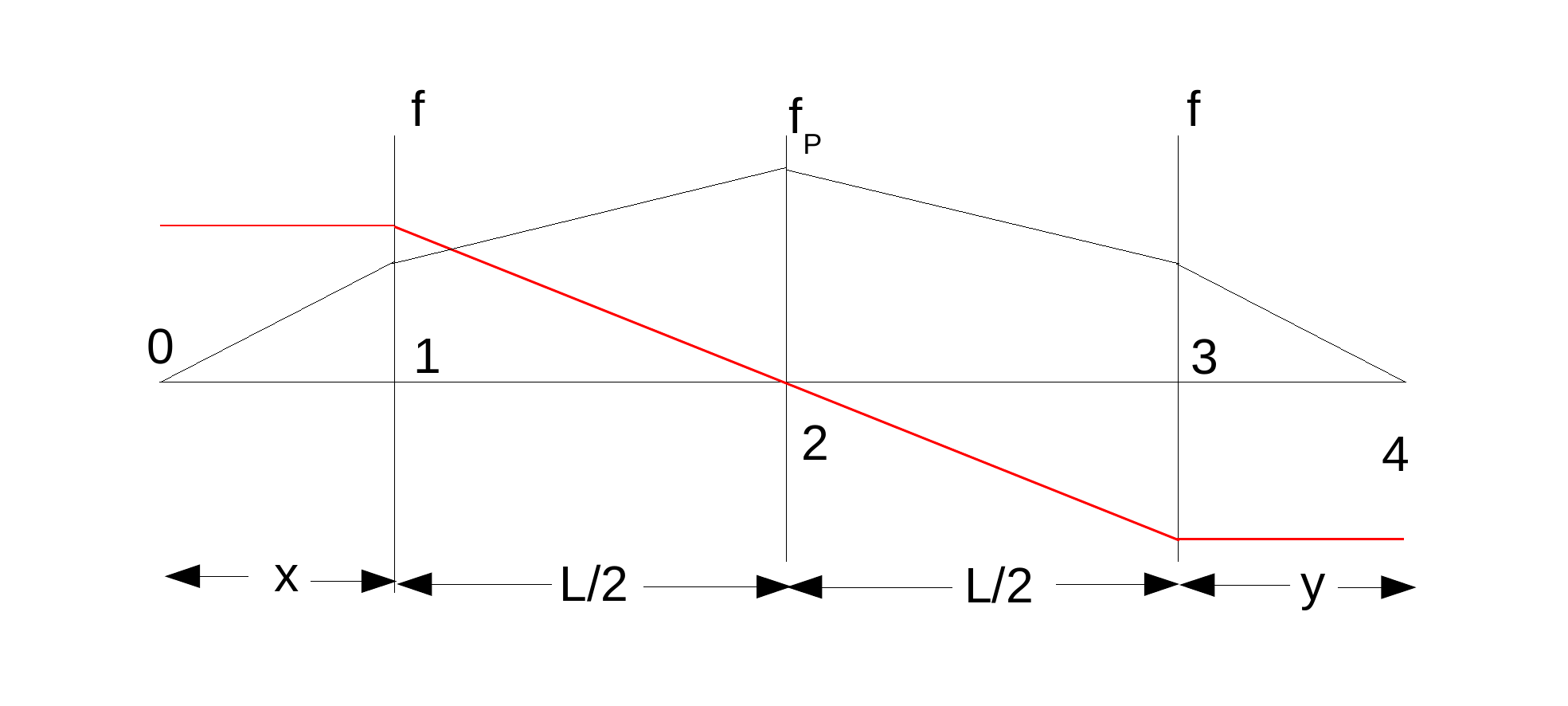}
  \end{center}
  \caption{\label{fig:pupil-relay} Schematic ray trace of a three-lens pupil relay system. Such a system can be used for beam transport or as the basis for a delay line. The design uses three lenses and requires a 10\,cm beam for $\lambda=1.65\,\micron$ and a 1\,km propagation distance. The first lens focuses the collimated input beam onto surface 2, and the third lens recollimates the beam (red ray). The second lens relays the telescope pupil image from surface 1 to surface 4 (black ray). The first and last lenses can double as the vacuum windows. }
\end{figure}

We will illustrate the general principles of pupil relay using simple series of thin lenses, which are straightforward to incorporate into a vacuum pipe. A three-lens design that relays the telescope pupil to a surface in the beam combining laboratory is presented and explained in Fig.~\ref{fig:pupil-relay}. For this design, the focal length $f_P$ of the pupil-remapping lens is $L/4$ in the limit $L \gg x$, $L \gg y$.

There are classes of designs with odd and even numbers of lenses. For the latter class, the overall loss from such a system can be minimised by replacing the vacuum windows at the ends of each pipe with lenses (of sufficient thickness to resist deformation due to the differential air pressure).

\begin{figure}
  \begin{center}
    \includegraphics[width=0.7\linewidth]{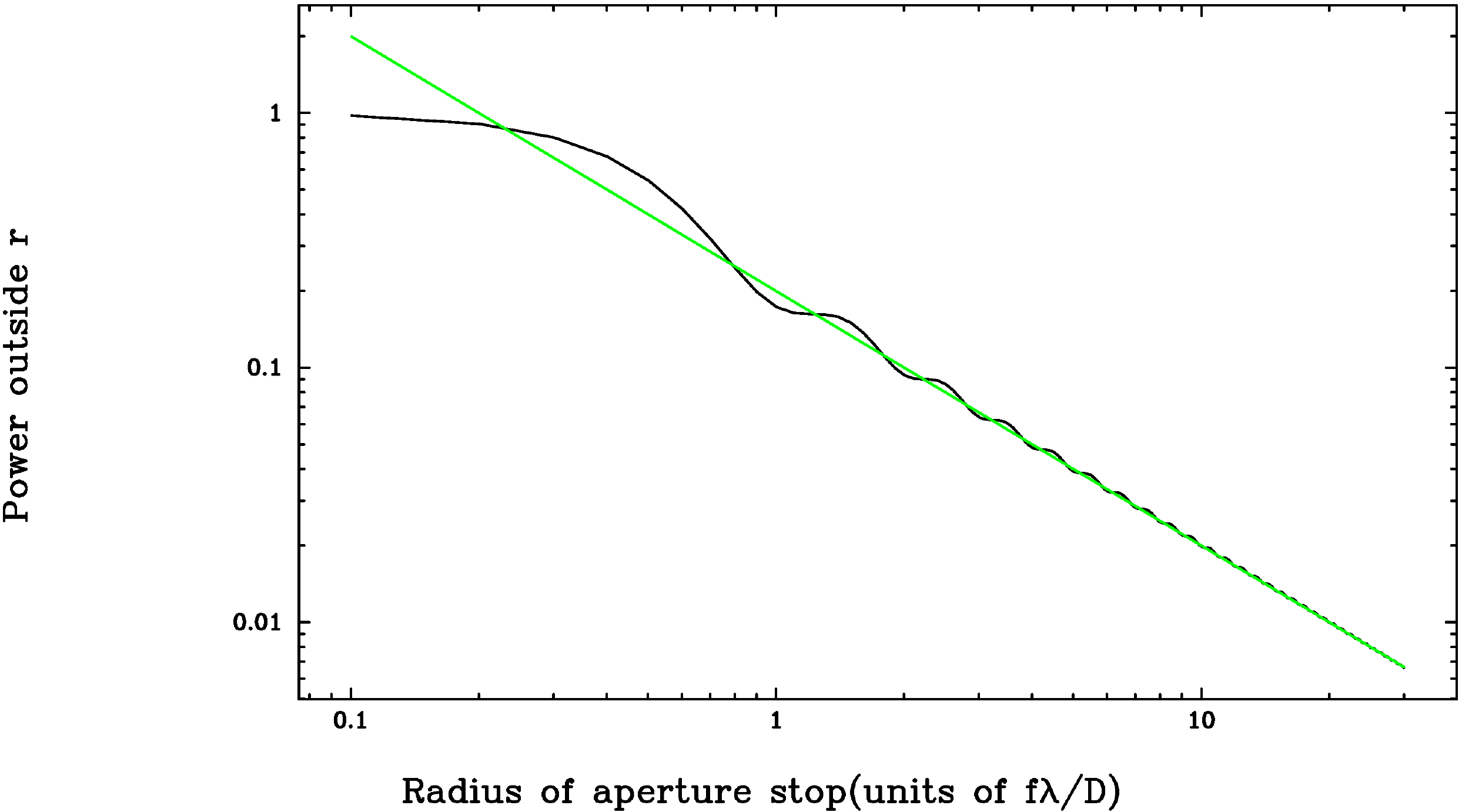}
  \end{center}
  \caption{\label{fig:truncate-Airy} Light loss caused by truncating an Airy disk. The green line is $\frac{\lambda f}{5AD}$ where $\lambda$ is the wavelength, $f$ the focal length, $D$ the diameter of the beam and $A$ the diameter of the aperture stop.}
\end{figure}

\begin{figure}
  \begin{center}
    \includegraphics[width=0.49\linewidth]{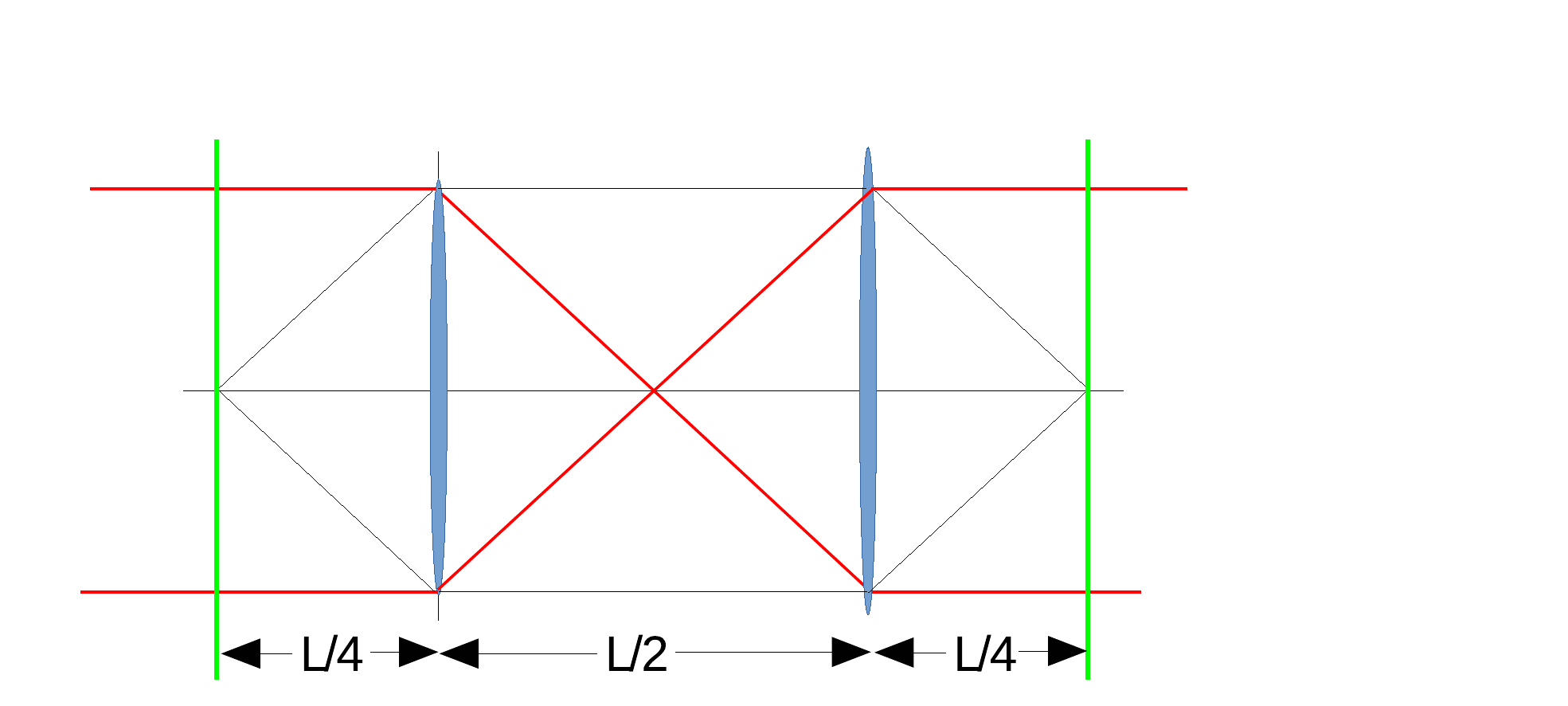}
    \includegraphics[width=0.49\linewidth]{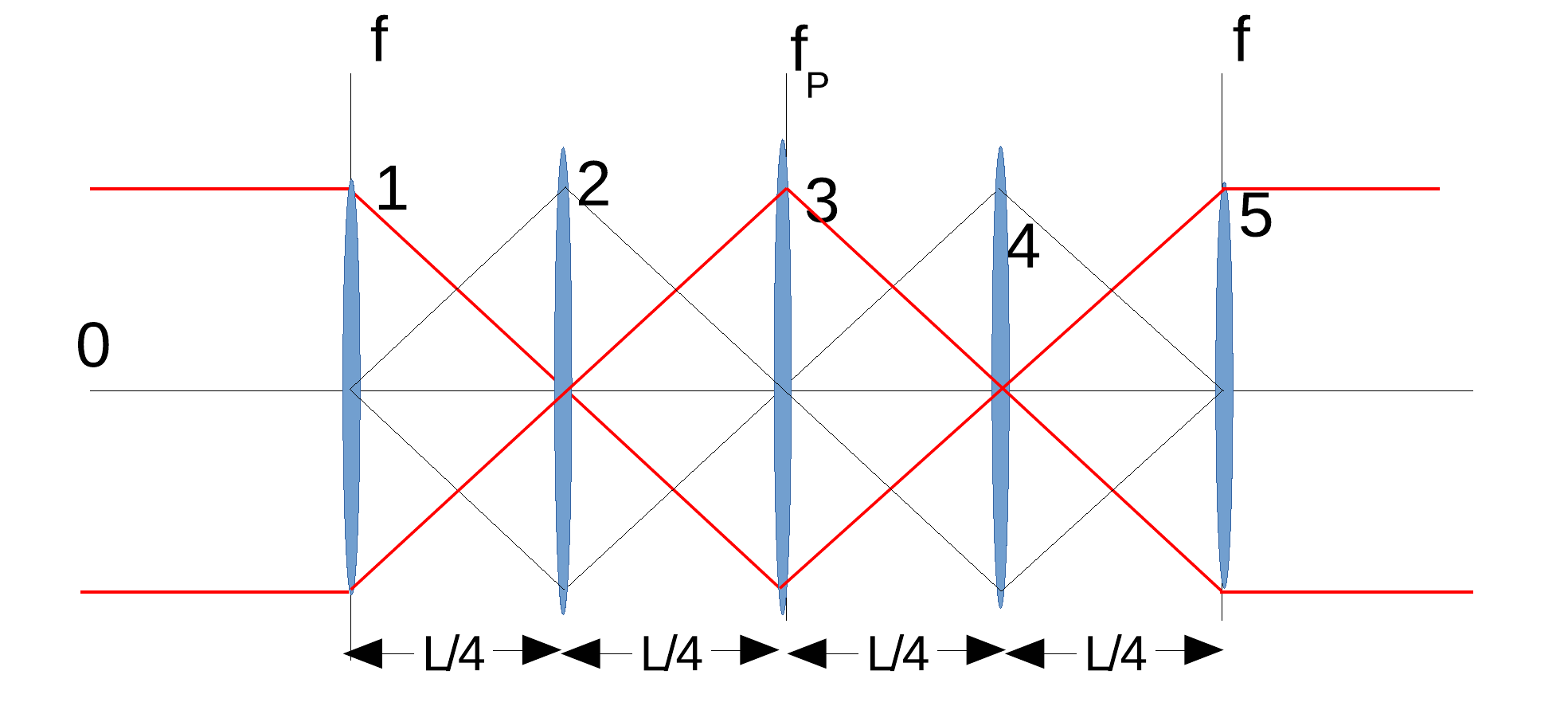}
  \end{center}
  \caption{\label{fig:thin-lens} Alternative pupil relay designs using different numbers of lenses. The left-hand design has the minimum number of powered surfaces (two). The right-hand design uses five lenses to reduce the beam diameter by 40\% compared with the three-lens system in \cref{fig:pupil-relay}. In both cases the pupil is transferred from the first to the last surface.}
\end{figure}

Given the initial beam diameter, the desired light loss imposes a minimum clear aperture requirement. In the absence of aberrations, an Airy disk will be formed at the pupil-remapping lenses (e.g.\ the middle surface in \cref{fig:thin-lens}), and the fractional light loss $l$ as a function of the aperture size $A$ there can be approximated (\cref{fig:truncate-Airy}) as:
\begin{equation}
l = \frac{2\lambda L}{5AD} \, .
\end{equation}

Since the optics will be in a vacuum pipe, we set $A=D$, hence:
\begin{equation}
D = \sqrt{\frac{2\lambda L}{5l}} \, .
\end{equation}
For 10\% light loss ($l=0.1$), $L=2$\,km and $\lambda=1.65\,\micron$, the required feed system diameter is just $D=11.5$\,cm. For $\lambda=13$\,\micron and $L=4$\,km we require $D=0.46$\,m, a factor of four smaller than that required for propagation without pupil relay.

\begin{figure}
  \begin{center}
    \includegraphics[width=0.8\linewidth]{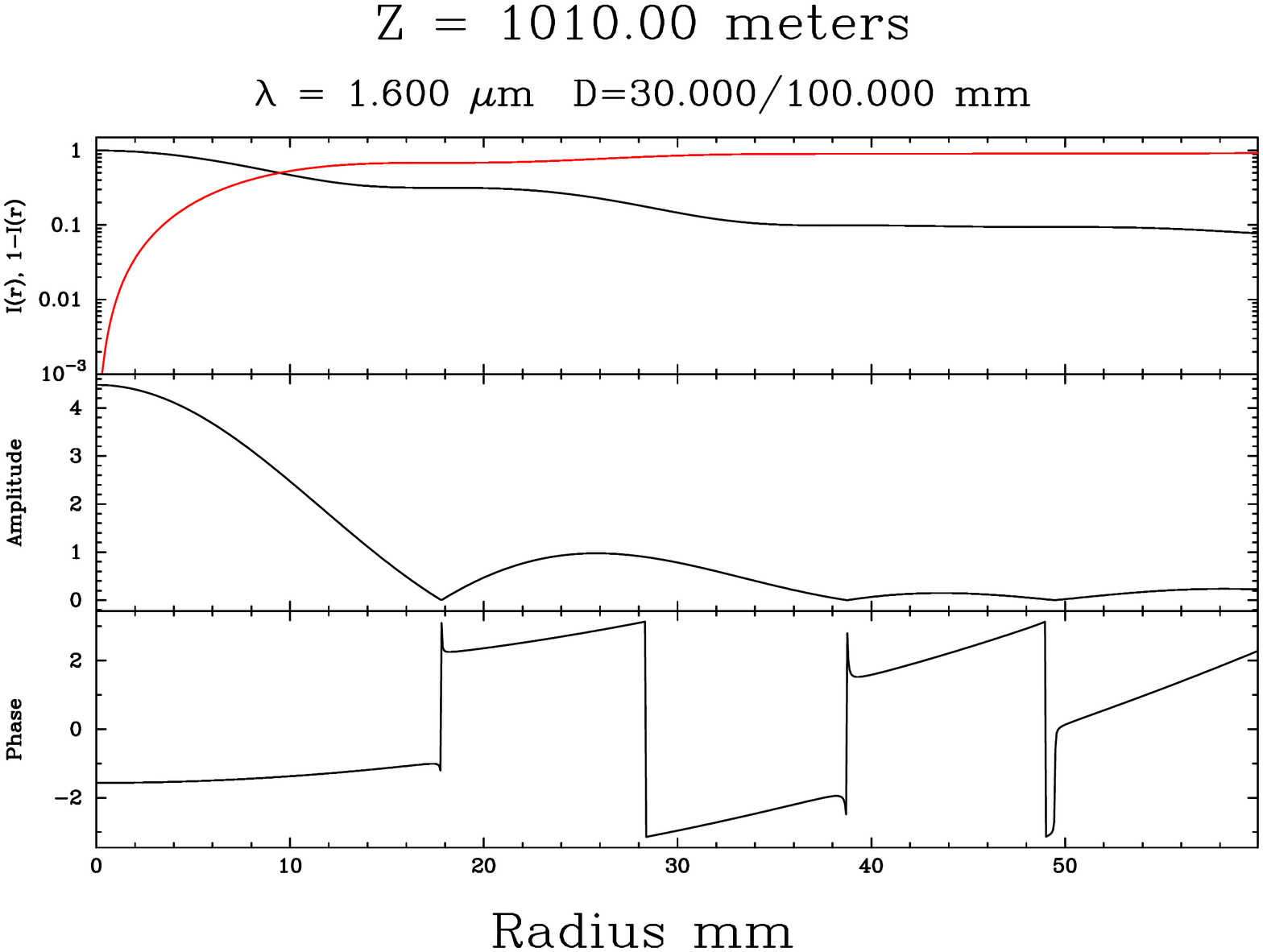}
  \end{center}
  \caption{\label{fig:beam-surf2} The beam profile at the middle of the feed system in \cref{fig:pupil-relay} (at the pupil-remapping lens). This is 1\,km after the entrance to the vacuum pipe. The top panel is the integrated flux within (red) or outside (black) the specified radius. The middle panel is the electric field amplitude (arbitrary units) and the bottom panel is the phase in radians. The entrance pupil is a perfect plane wave with a diameter of 100\,mm and a 30\,mm central obscuration. All surfaces have 120\,mm diameter stops.}
\end{figure}

\begin{figure}
  \begin{center}
    \includegraphics[width=0.8\linewidth]{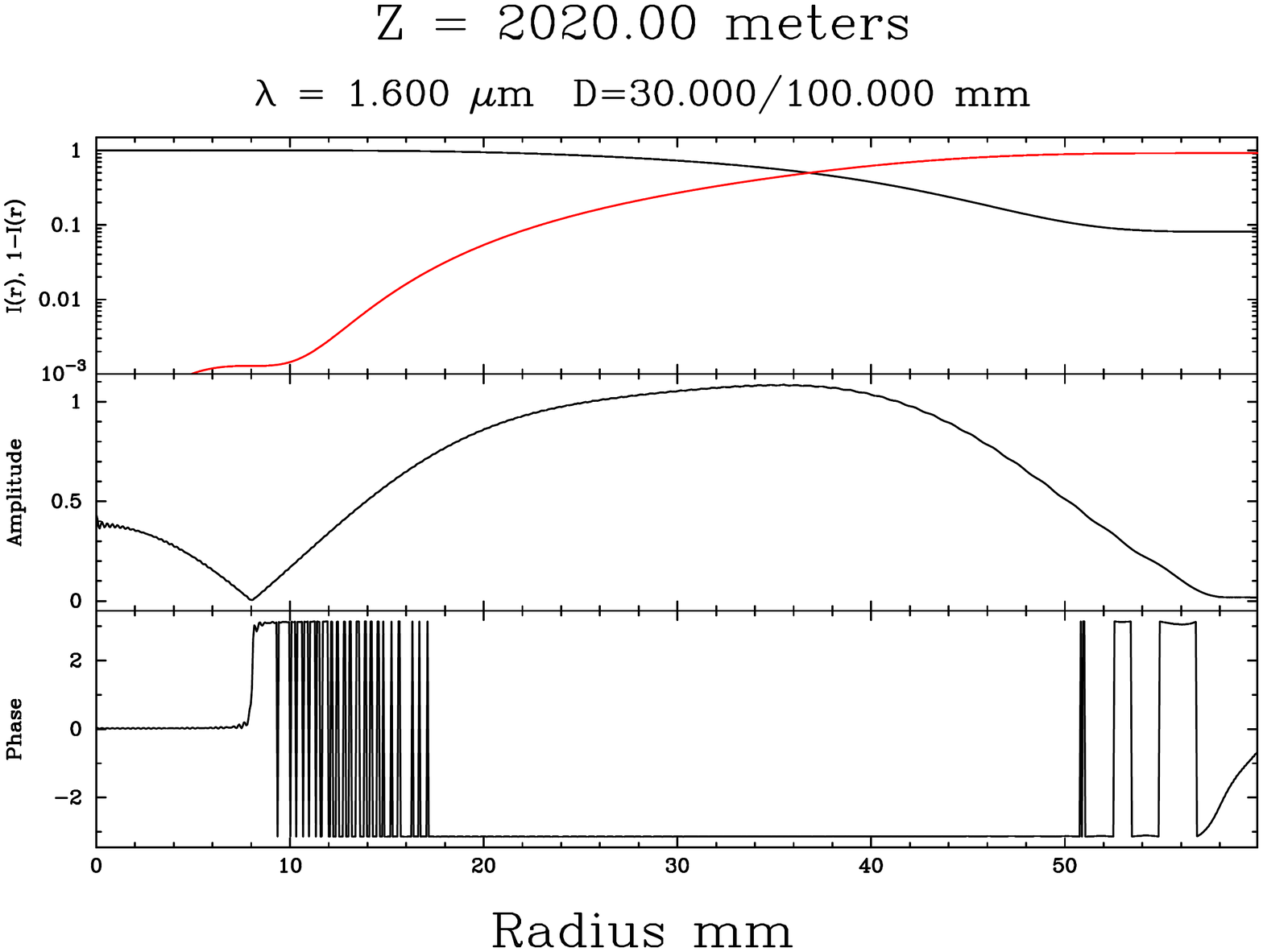}
  \end{center}
  \caption{\label{fig:beam-surf4} The beam profile at the exit pupil of the feed system in \cref{fig:pupil-relay}, 10\,m past the end of the vacuum pipe.}
\end{figure}

To illustrate the propagation effects, we have evaluated the Fresnel integrals numerically to determine the field at several surfaces of interest, assuming a perfect plane wave ($D=0.1$\,m) with a 30\,mm central obscuration at the entrance pupil (surface 0 in \cref{fig:pupil-relay}). \cref{fig:beam-surf2,fig:beam-surf4} show the $\lambda=1.65\,\micron$ beam profiles in the middle of the vacuum system (surface 2) and at the exit pupil (surface 4), for a system with $L=2$\,km.

It is important to match the beam profiles to each other, not to the input profile. When light is transported a shorter distance, light loss is decreased and the beam profiles will no longer match causing a reduction in fringe visibility. The profiles can be matched by changing the size of the stops at the pupil-remapping lens.

The beam diameter can be reduced by adding more lenses. \cref{fig:thin-lens} shows a design with five lenses, as well as a design with just two lenses. The latter is part of the family of designs using even numbers of lenses.

For all of the designs considered here, the lenses must deliver good image quality across both the near-IR fringe tracking band and the mid-IR science band(s), but need not at the intervening wavelengths. The focal length must also not vary too much with wavelength. We have made an initial exploration of suitable lens designs, and it appears that doublet designs can exceed the required focus tolerance by an order of magnitude.

Reflective equivalents of these designs are of course perfectly valid. In the next section we consider these for use in a system of switchable fixed delays.

\section{Delay line concepts}
\label{sec:dl-concepts}

\subsection{Variable delay line}

\begin{sidewaysfigure}
  \includegraphics[width=\linewidth]{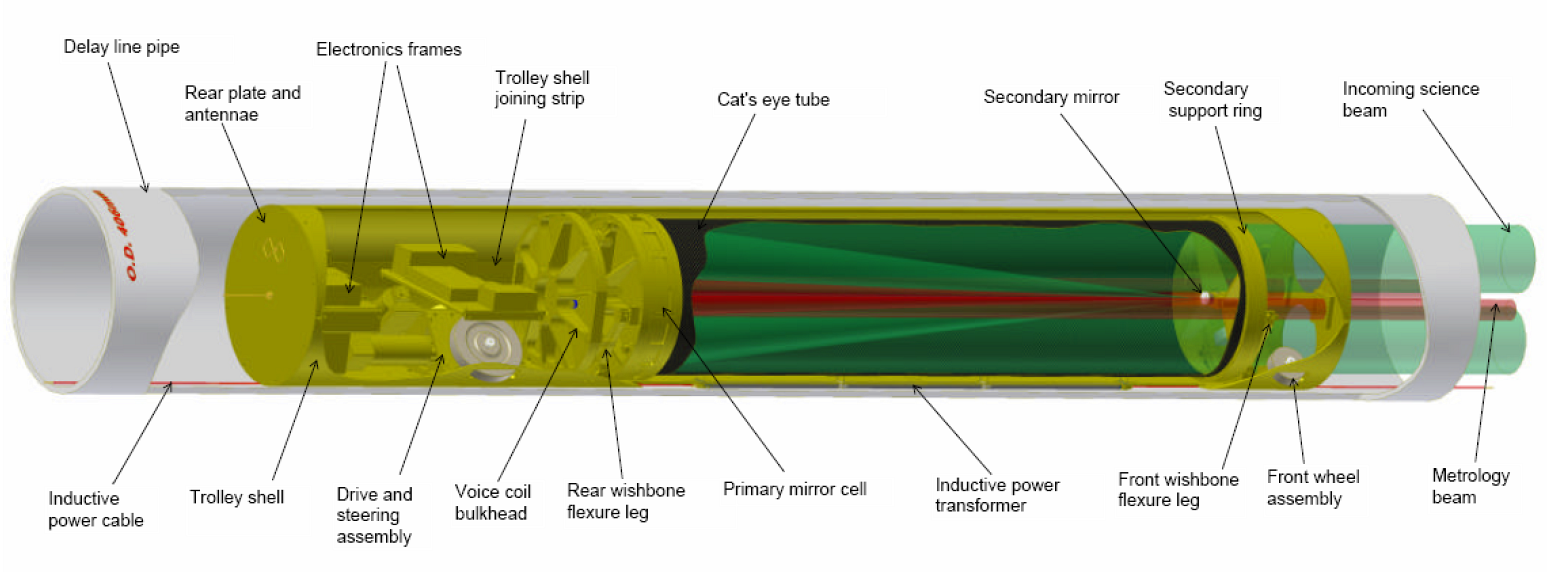}
  \bigskip
  \includegraphics[width=0.89\linewidth]{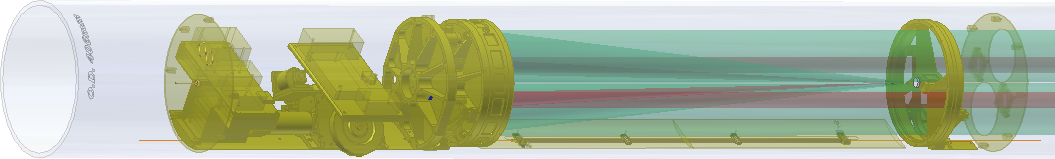}
  \bigskip
  \caption{\label{fig:MROI_DL} Diagram of the MROI delay line trolley, showing the physical locations and space envelope of the active elements. The top part of the diagram shows the complete trolley (comprising the cat's-eye retro-reflector and the motorized carriage) inside the vacuum pipe, and the lower part shows only the cat's eye. The diameter of the carriage tube is approximately 14 inches, to accommodate a clear aperture of 125\,mm for the input and output beams. The approximate wheelbase of the carriage is 1.8\,m.}
\end{sidewaysfigure}

The longest continuously-variable vacuum delay line in existence is at the Magdalena Ridge Observatory Interferometer (MROI). The MROI delay lines \cite{2010SPIE.7734E..49F} have been designed to provide zero to 380\,m of continuously-variable delay in vacuum, and have been fully tested up to 200\,m of delay. The optical design is a straightforward cat's-eye without pupil relay. The innovative features are the use of the inner surface of the vacuum pipe to guide and support the movable ``trolley'' (\cref{fig:MROI_DL}) that carries the retro-reflecting optics, in conjunction with active control of the return beam shear (to accommodate non-straightness of the vacuum pipe) and the trolley roll angle. To avoid issues with dragging long cables, power is supplied inductively and all communications with the trolley are wireless. The factors limiting straightforward extension of the design to a longer stroke are the coherence length of the laser used in the off-the-shelf metrology system ($\sim$1\,km) and the 1.4\,m/s (in delay) slew speed (limited by the inductive power supply and onboard energy storage).

A further option for extending the range of the delay line is to send the beam through the delay line multiple times. This has the advantage of magnifying the slew speed for faster repositioning. Using multiple passes obviously reduces the allowed jitter of the retro-reflecting optics by the reciprocal of the number of passes. MROI has been designed for visible and near-infrared operation, and hence a requirement that the induced optical path jitter not exceed $\lambda/40$ at $\lambda=600$\,nm. The design is thus over-specified for the shortest PFI wavelength of 1.65\,\micron, and this margin could be used to accommodate several passes while still delivering satisfactory performance.

The MROI delay line design has a clear aperture of 125\,mm for the optical beam using vacuum pipe of 15.5 inch (394\,mm) inner diameter. The diffraction calculations presented in \cref{fig:propagation} indicate that scaling up the dimensions by a factor $\sim$2 would be adequate for fringe tracking at 1.65\,\micron wavelength on baselines up to 2\,km. For science observations in the mid-IR, it would be necessary to use switchable fixed delays in conjunction with short delay lines, and/or incorporate pupil management into the delay line optics. This could be done using a variable curvature mirror as implemented at VLTI \cite{2000SPIE.4006..104F}. A possible scheme for combining a pupil-managed variable delay line with beam transport from the telescopes to the central beam combining laboratory is shown in \cref{fig:shared-pipe} and discussed below.

\subsection{Switchable fixed delays}

We can use the pupil relay schemes presented in \cref{sec:pupil-relay} as the basis of designs for ``long delay lines'' that introduce switchable fixed delays in vacuum.

\begin{figure}
  \begin{center}
    \includegraphics[width=\linewidth]{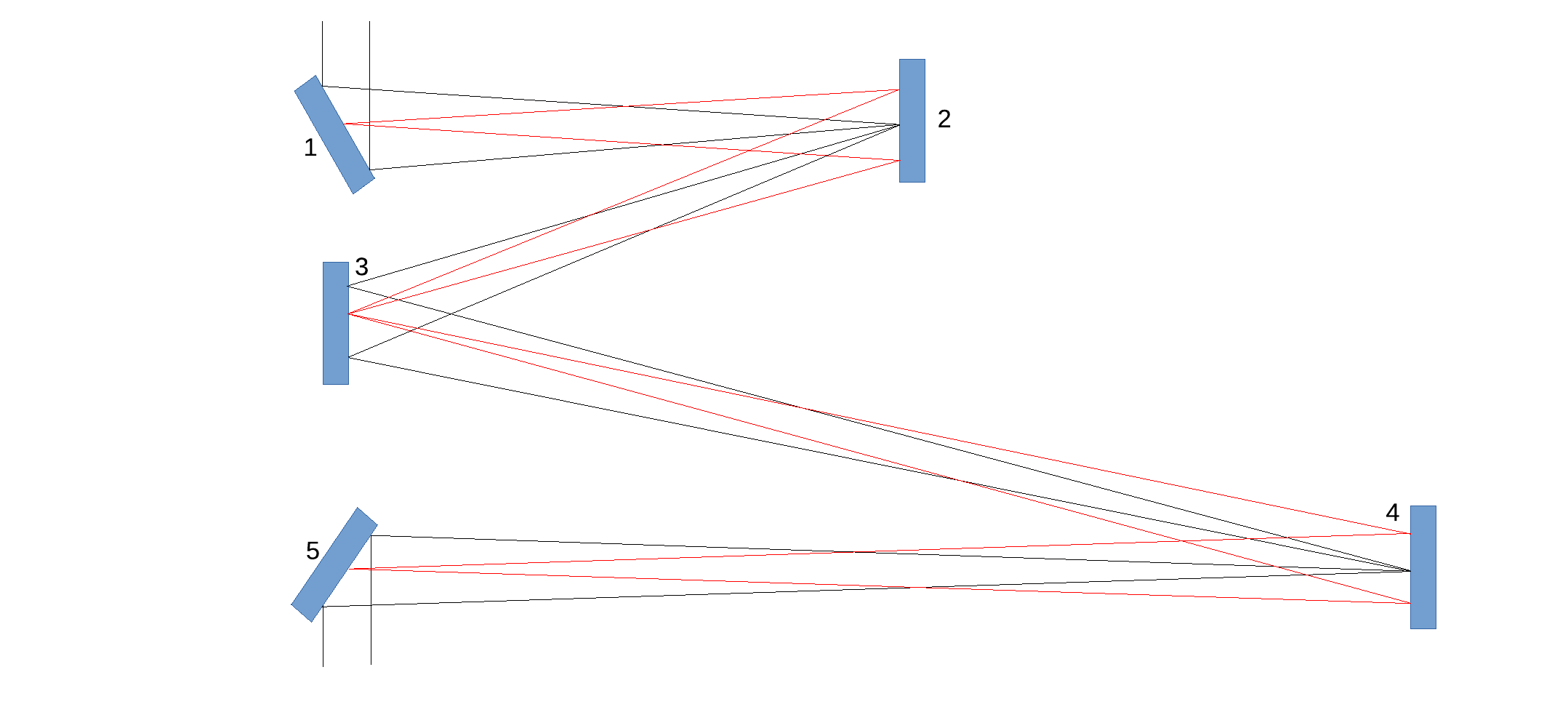}
  \end{center}
  \caption{\label{fig:w-design} Asymmetric W-shaped long delay line concept (not to scale). The collimated beam enters at the top of the diagram and is delayed by being reflected off surfaces 1 to 5 in sequence. Mirror 1 focuses the beam onto mirror 2. Mirror 3 transfers an image of mirror 2 to mirror 4 and mirror 5 recollimates the beam (black rays). The pupil is relayed from mirror 1 to mirror 5 as follows: mirror 2 makes an image of mirror 1 on mirror 3 and mirror 4 images mirror 3 to mirror 5 (red rays). The horizontal axis is greatly compressed --- multiple copies of mirrors 2 and 4 are located at various positions along a vacuum pipe. Mirrors 1, 3, and 5 must have variable curvature to accommodate the various possible locations of mirrors 2 and 4.}
\end{figure}

A possible scheme for such a ``long delay line'' is shown and explained in \cref{fig:w-design}. This ``asymmetric W'' design is equivalent to the five-lens design shown in \cref{fig:thin-lens}, and requires three variable-curvature surfaces to accommodate a discrete set of positions for the other two surfaces (mirrors 2 and 4), thus allowing appropriate fixed delays to be introduced. Fast switching of the fixed delays would probably be accomplished by making mirrors 2 and 4 ``pop-up'' mirrors that fold down below the beam when not in use.

A simplification of the asymmetric W can be realised by adopting the two-lens pupil relay scheme from \cref{fig:thin-lens}. In this scheme, only the pop-up mirrors have power, and the three other surfaces are just flats, at the cost of larger optics.

\begin{figure}
  \begin{center}
    \includegraphics[width=0.69\linewidth]{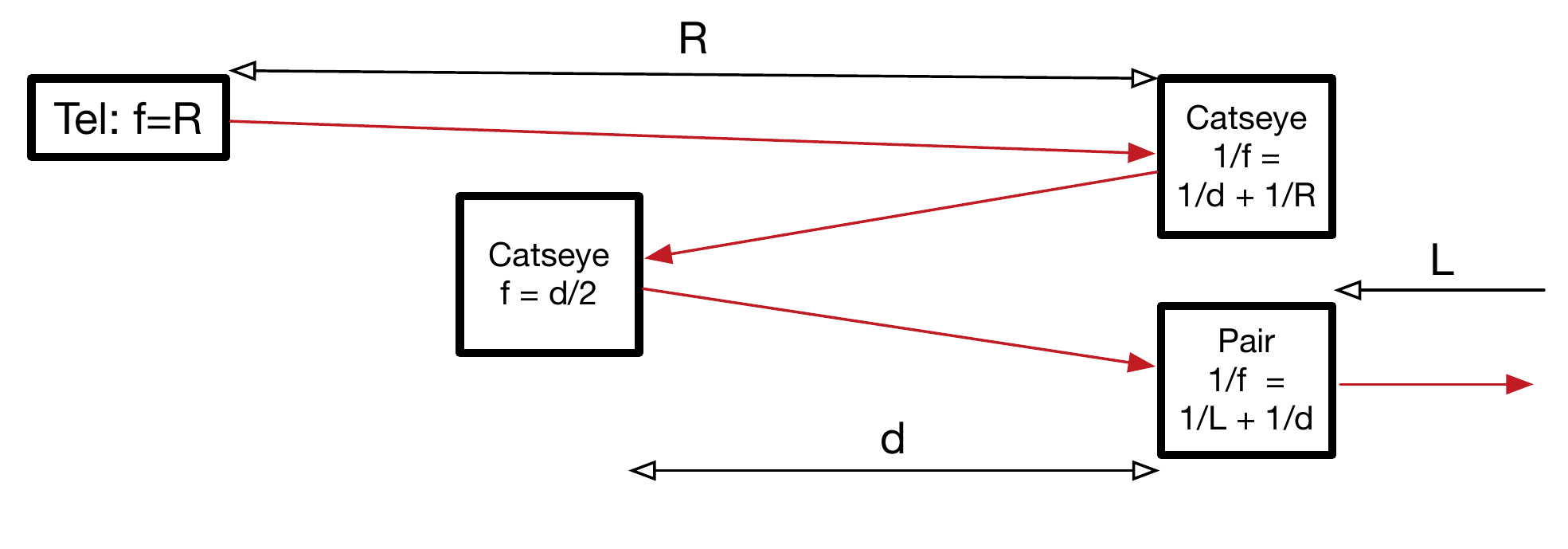}
    \includegraphics[width=0.25\linewidth]{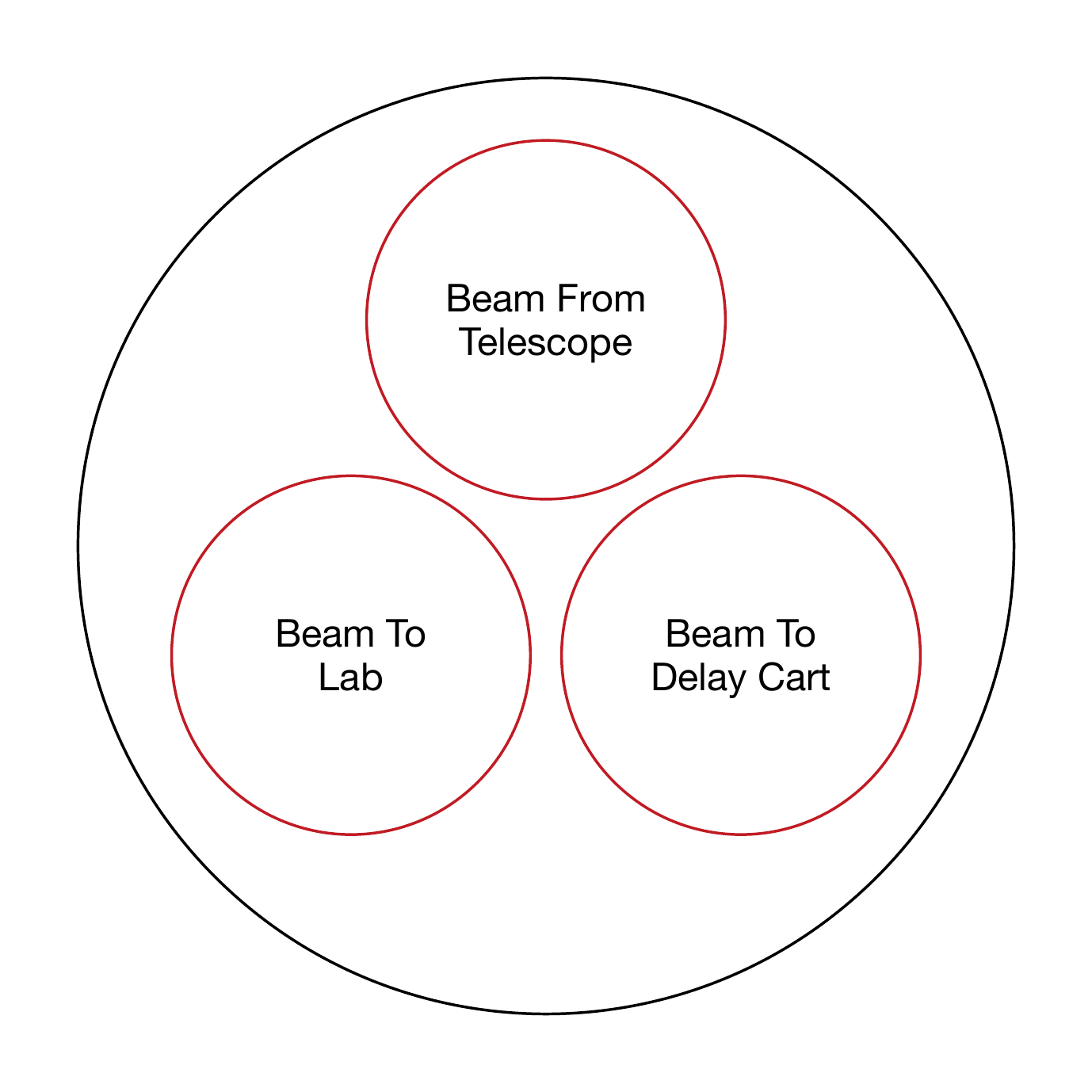}
  \end{center}
  \caption{\label{fig:shared-pipe} A schematic of an example beam transport architecture for PFI, with diffraction losses below 10\%. At no point are beams collimated: the telescope forms a focus at the array center, a distance $R$ from the telescope. A fixed-position cat's-eye forms a telescope pupil image on a delay line cat's-eye a variable distance $d$ away. This image is relayed by ``pair'' (which could be implemented as a pair of confocal paraboloids) to a surface a distance $L$ away in the beam combining laboratory. The last three components require variable curvature. The delay line and transport from the telescope could share the same 0.6\,m vacuum pipe, as shown in the right-hand panel.}
\end{figure}

To further reduce the amount of real estate and vacuum pipe required, the beam transport and delay line functions can be combined in a single vacuum pipe. Fig.~\ref{fig:shared-pipe} shows a concept design that uses variable curvature mirrors (or multiple curved surfaces of adjustable geometry) to realize this idea.

\section{Conclusions}
\label{sec:conclusions}

We have evaluated the likely diffraction losses in the PFI beam transport and delay line systems. For direct detection interferometry in the mid-infrared on kilometric baselines, it is clear that these systems must incorporate pupil management to avoid the need for prohibitively large (1--2\,m diameter) and hence expensive vacuum pipes. We have presented conceptual designs for switchable fixed delays and continuously-variable delays that incorporate pupil relay. No show-stopping issues were identified.

As we continue to develop the technology roadmap for PFI, we plan to evaluate the possibilities for using optical fibres as an alternative to free-space propagation in vacuum, and to refine the design ideas presented in this paper with a view to selecting the most viable concepts.

\bibliography{PFI_beam_trans_manuscript}
\bibliographystyle{spiebib}

\end{document}